\documentstyle[epsfig,aps,prb]{revtex}
\newcommand\beq{\begin{equation}}
\newcommand\eeq{\end{equation}}
\newcommand\bea{\begin{eqnarray}}
\newcommand\eea{\end{eqnarray}}
\newcommand\aial{{\vec {\cal A}}_{i, \alpha} }
\newcommand\rial{{\vec r}_{i, \alpha} }
\newcommand\rjbe{{\vec r}_{j, \beta} }
\newcommand\xiab{\xi_{\alpha \beta} }

\begin{document}

\begin{center}
{\Large A Solvable Model of Interacting Fermions in Two Dimensions}
\end{center}

\vskip .5 true cm
\centerline{\bf B. Sriram Shastry \footnote{E-mail address: 
bss@physics.iisc.ernet.in}}
\centerline{\it Department of Physics,}  
\centerline{\it Indian Institute of Science, Bangalore 560012, India} 
\vskip .5 true cm

\centerline{\bf Diptiman Sen \footnote{E-mail address: 
diptiman@cts.iisc.ernet.in}} 
\centerline{\it Centre for Theoretical Studies,}  
\centerline{\it Indian Institute of Science, Bangalore 560012, India} 
\vskip .5 true cm

\begin{abstract}

We introduce and study an exactly solvable model of several species of 
fermions in which particles interact pairwise through a mutual magnetic field; 
the interaction operates only between particles belonging to different 
species. After an unitary transformation, the model reduces to one in which 
each particle sees a magnetic field which depends on the total numbers of 
particles of all the other species; this may be viewed as the mean-field model 
for a class of anyonic theories. Our model is invariant under charge 
conjugation $C$ and the product $PT$ (parity and time reversal). For 
the special case of two species, we examine various properties of this system, 
such as the Hall conductivity, the wave function overlap arising from the 
transfer of one particle from one species to another, and the one-particle 
off-diagonal density matrix. Our model is a generalization of a recently 
introduced solvable model in one dimension.

\end{abstract}
\vskip .5 true cm

~~~~~~ PACS number: ~71.10.Pm, 71.27.+a

\vskip 1 true cm

Exactly solvable models of interacting particles have often been very useful
in illustrating some general concepts in many-body physics. While there is a 
large variety of such models available in one dimension, many of which fall 
into the class of Tomonaga-Luttinger liquids \cite{tomo}, there are few 
models known in two dimensions which are completely solvable. In this paper,
we introduce and study a model of several species of fermions which 
interact with each other through a magnetic field term which depends on
the coordinates of pairs of particles belonging to two different species. The
model can be solved by a unitary transformation which reduces it
to a model of fermions in a magnetic field which depends on the total numbers 
of fermions belonging to the other species. Our model is a direct 
generalization of the recent reinterpretation of the well-known model of
Luttinger in one dimension \cite{schu1}. The one-dimensional model also has 
pairwise ``gauge" interactions depending on the coordinates of the particles; 
the model is exactly solvable because the interactions can be 
unitarily gauged away at the cost of modifying the boundary conditions in
a non-trivial way. As we will see, in our two-dimensional model the 
interactions cannot be gauged away in the bulk of the system; the unitary 
transformation leaves behind a static magnetic field.

Let us consider $\nu$ species of fermions in two dimensions (say, the
${\hat x}-{\hat y}$ plane), with the charge and number of fermions of type 
$\alpha$ being denoted by $q_\alpha$ and $N_\alpha$ respectively. The 
coordinates of the particles will be denoted by $\rial$, where $1 \le i 
\le N_\alpha$ and $1 \le \alpha \le \nu$. We will consider the Hamiltonian
\bea
{\cal H} ~&=&~ \sum_{i, \alpha} ~\frac{1}{2m_\alpha} ~\Bigl( ~{\vec p}_{i, 
\alpha} ~-~ \frac{q_\alpha}{c} {\vec A}_{i, \alpha} ~-~ \frac{q_\alpha}{c} 
\aial ~\Bigr)^2 ~, \nonumber \\
{\vec A}_{i, \alpha} ~&=&~ \frac{1}{2} ~B_0 ~{\hat z} ~\times ~\rial ~,
\label{ham}
\eea
where $c$ is the velocity of light. $B_0 {\hat z}$ is an external magnetic
field pointing in a direction perpendicular to the two-dimensional plane; 
we have chosen the symmetric gauge for its vector potential ${\vec A}_{i, 
\alpha}$ in order to explicitly maintain invariance under rotations of the
plane. The other vector potential $\aial$ arises from two-body interactions; 
it will be taken to have the following form which is natural in two 
dimensions \cite{schu2},
\beq
\aial ~=~ \eta_\alpha ~\sum_{j \beta} ~\xiab ~{\hat z} ~ \times ~(~ \rial ~-~ 
\rjbe ~)~,
\label{gauge}
\eeq
where $\eta_\alpha$ and $\xiab$ are some constants to be fixed below. Note 
that the Hamiltonian (\ref{ham}) is invariant under translations in the plane.

We may now perform an unitary transformation on the Hamiltonian of the form
\bea
{\tilde {\cal H}} ~&=&~ U ~{\cal H} ~U^{-1} ~, \nonumber \\
U ~&=&~ \exp ~\Bigl[ ~-~ \frac{iq}{\hbar c} ~\sum_{\alpha < \beta} 
\sum_{i,j} ~\xiab ~{\hat z} ~\cdot ~\rial ~\times \rjbe ~\Bigr] ~.
\label{unit}
\eea
where $q$ is the charge of an electron.
(We note that the phase factor in $U$ only depends on the total coordinates 
${\vec R}_\alpha = \sum_i {\vec r}_{i, \alpha}$ of the various species of 
fermions). This gives the transformed Hamiltonian
\bea
{\tilde {\cal H}} ~&=&~ \sum_{i,\alpha} ~\frac{1}{2m_\alpha} ~\Bigl( ~{\vec 
p}_{i, \alpha} ~-~ \frac{q_\alpha}{c} {\vec A}_{i, \alpha} ~-~ 
\frac{q_\alpha}{c} {\vec a}_{i, \alpha} ~ \Bigr)^2 ~, \nonumber \\
{\vec a}_{i, \alpha} ~&=&~ \frac{1}{2} ~\Bigl( ~\eta_\alpha ~
\sum_{\beta \ne \alpha} ~\xiab ~N_\beta ~\Bigr) ~{\hat z} ~\times ~\rial ~, 
\label{hamtil}
\eea
provided that 
\bea
\xiab ~&=&~ - ~\xi_{\beta \alpha} ~, \nonumber \\
{\rm and} \quad q_\alpha \eta_\alpha ~&=&~ q \quad {\rm for ~~all ~~} 
\alpha ~.
\eea
The antisymmetry of $\xi_{\alpha \beta}$ implies that the two-particle 
magnetic interaction can only act between particles belonging to two 
different species. 

It is interesting to consider the effects of some discrete symmetries such 
as time reversal ($T$), parity ($P$) and charge conjugation ($C$). Let us 
first set the external magnetic field $B_0 = 0$. Under $T$, the wave 
functions and factors of $i$ are complex conjugated (thus, the 
momentum operators ${\vec p}_{i, \alpha}
\rightarrow - {\vec p}_{i, \alpha}$) and the time coordinate $t \rightarrow - 
t$; the space coordinates $x,y$ and the various parameters $q_{\alpha}, 
\eta_{\alpha}$ and $\xiab$ remain unchanged. 
Under $P$, one of the space coordinates, say, $x \rightarrow -x$, 
while $y$, $t$ and all the parameters remain unchanged. We therefore see
that the model is not invariant under $P$ and $T$ separately, but it is
invariant under the combined operation $PT$. Under charge conjugation, we 
demand that $q_{\alpha} \rightarrow - q_{\alpha}$ and $\xiab \rightarrow -
\xiab$, while $\eta_{\alpha}$ and the space-time coordinates remain unchanged; 
thus the model is invariant under $C$ and therefore under $CPT$. Finally, if 
the external magnetic field $B_0$ is nonzero, the model is again invariant 
under $C$ and $PT$, but not under $P$ and $T$ separately; this is because a 
magnetic field (which must be produced by some external currents) changes sign 
under $C$, $P$ and $T$ separately.

It may be useful to point out here that our model has some resemblance to the 
mean field theory of several species of anyons. In the usual theories of 
anyons, the wave function is assumed to pick up a phase $\theta_{ij}$ whenever 
particle $i$ is taken in an anticlockwise loop around particle $j$, no matter 
what the size and shape of the loop is \cite{lein}. This is
often modeled by treating each particle as a point-like composite of 
charge and magnetic flux; when one particles encircles another, the wave
function picks up an Aharonov-Bohm phase. In understanding the many-body 
properties of such a system, a fruitful approach has been to begin with a 
mean field theory in which the magnetic flux of each anyon is smeared out over 
the entire plane \cite{fett,chen}. Thus each particle sees a magnetic 
field proportional to the average density of particles, which is similar to 
our situation. Of course, the analysis of anyons then goes beyond mean field
theory to study the fluctuations about the average magnetic field, while our
simplified model has no fluctuations. It is worth remarking that our model has
no counterpart for the most popular anyon model which has only one species;
we need a a minimum of two species.

To continue, the total magnetic field seen by a particle of type $\alpha$ in
our model is given by $B_\alpha {\hat z}$, where
\beq
B_\alpha ~=~ B_0 ~+~ \eta_\alpha ~\sum_{\beta \ne \alpha} ~\xiab ~
N_\beta ~.
\label{btot}
\eeq
In order to have a well-defined thermodynamic limit $N_{\alpha} \rightarrow
\infty$, the $\xiab$ must be taken to scale as $1/A$, where $A$ is the area
of the system; thus the magnetic field strengths $B_\alpha$ in (\ref{btot})
remain of order $1$ as $A \rightarrow \infty$ with the densities 
$\rho_\alpha = N_\alpha /A$ held fixed. We then expect Landau levels to 
form for each species \cite{land}. It is well-known that each Landau level has 
a macroscopic degeneracy equal to $A\vert q_{\alpha} B_{\alpha} \vert /(2\pi 
\hbar c)$. The filling fraction of fermions of type $\alpha$ is given by 
\beq
f_\alpha ~=~ \rho_\alpha ~\frac{2\pi \hbar c}{\vert q_\alpha B_\alpha \vert}~.
\eeq
If $f_\alpha$ is not equal to an integer for one or more values of $\alpha$, 
the ground state of the system is highly degenerate. 

For computational purposes, it is convenient to 
break this degeneracy in one of two ways. We can either add a simple harmonic 
confining potential to the Hamiltonians (\ref{ham}) and (\ref{hamtil}) of the 
form
\beq
{\cal H}_{sh} ~=~ \frac{k}{2} ~\sum_{i, \alpha} ~{\vec r}_{i, \alpha}^2 ~,
\eeq
and take the limit $k \rightarrow 0$ at the end of the calculation, or we can 
simply impose a hard wall boundary condition at some large radius $R$. 
Analytically, it is easier to work with the first method since the problem of 
free particles in a combination of an uniform magnetic field and a simple 
harmonic confinement is exactly solvable as we will now discuss. (Let us drop 
the species label $\alpha$ in the rest of this paragraph and in the next). 
Since the problem has rotational symmetry, the energies and wave functions are 
specified by two quantum numbers, a radial quantum number $n=0,1,2,...$ and 
the angular momentum $l=0, \pm 1, \pm 2, ...$. If only a magnetic field is
present (with, say, the product $qB$ being positive), the 
single-particle states have energies which only depend on the integer 
$n$ which counts the number of nodes in the radial direction; thus
\bea
E_{n,l} ~&=&~ \hbar \omega_c ~(~ n ~+~ \frac{1}{2} ~)~, \nonumber \\
\omega_c ~&=&~ \frac{qB}{mc} ~.
\eea
In the lowest Landau level (LLL), $n=0$ while $l$ can only take non-negative
values; all states have the energy $E_{0,l}= \hbar \omega_c /2$ independent 
of $l$. The normalized wave functions in the LLL are given in 
terms of the complex coordinates $z=x+iy$ and $z^\star = x-iy$ as
\beq
\psi_{0,l} (z, z^\star ) ~=~ \Bigl( ~\frac{qB}{2\hbar c} ~\Bigr)^{(l+1)/2} ~
\frac{z^l}{\sqrt {l! ~\pi}} ~\exp \Bigl[ ~-~ \frac{qB}{4\hbar c} z z^\star ~
\Bigr] ~,
\label{wave}
\eeq
where $l=0,1,2,...$. The amplitudes of these 
wave functions are peaked on circles of various radii centered about the 
origin ${\vec r} = {\vec 0}$; the radii of these ``ring" states are given by 
$r_l = \sqrt{2l \hbar c/(qB)}$. (If $qB$ is negative, the LLL wave functions
are given by Eq. (\ref{wave}) with $z$ replaced by $z^\star$. Then the
angular momentum only takes non-positive values).

If we now add a weak simple harmonic potential $m\omega^2 {\vec r}^2 /2$ for 
all the particles, the energies of the ring states in the LLL become
\beq
E_l ~=~ \frac{\hbar}{2} ~\Bigl[ ~\sqrt{\omega_c^2 ~+~ 4 \omega^2} ~+~ (~
\sqrt{\omega_c^2 ~+~ 4 \omega^2} ~-~ \omega_c ~) ~\vert l \vert ~\Bigr] ~,
\eeq
which increase from the origin outwards as $\vert l \vert$ increases from
zero. In the 
many-particle ground state, therefore, the fermions fill up the individual 
ring states from the origin outwards. In the following discussion, we will
assume this order of filling in the LLL, without explicitly 
mentioning the simple harmonic confinement which justifies it.

We will now specialize to the case of two species of fermions to illustrate 
some properties of our model. Let us take the masses equal to $m$ for both
species, the charges equal to $q_1 =q_2 = q$ (thus, $\eta_1 = \eta_2 =1$), 
and the numbers of particles equal to $N_1$ and $N_2$ for the two species 
respectively. We will also set 
\beq
\xi_{12} ~=~ - ~\xi_{21} ~=~ \frac{\gamma}{A} ~,
\label{gam}
\eeq
where $\gamma$ is a number of order $1$. After the unitary transformation in 
(\ref{unit}), the two species see uniform magnetic fields equal to
\bea
B_1 ~&=&~ B_0 ~+~ \gamma ~\frac{N_2}{A}  \nonumber \\
{\rm and} \quad B_2 ~&=&~ B_0 ~-~ \gamma ~\frac{N_1}{A} 
\label{mag}
\eea
respectively. If the number of particles $N_1 = N_2$, the model is invariant
under the exchange of the species labels
$1 \leftrightarrow 2$ and $\gamma \rightarrow - \gamma$; this is in addition
to the discrete symmetries $C$ and $PT$ discussed in general before.

One of the properties of interest for such a model is the Hall conductivity.
In the absence of impurities and any other interactions (such as Coulomb
repulsion), what is the Hall conductivity of this system if the filling 
fractions $f_1$ and $f_2$ are both integers? It is fairly easy to see that
the answer is 
\beq
\sigma_{xy} ~=~ [~ f_1 ~{\rm sign} (B_1) ~+~ f_2 ~{\rm sign} (B_2) ~] ~
\frac{q^2}{2\pi \hbar} ~.
\label{sigxy}
\eeq
This can be derived from the usual formula for the frequency-dependent
conductivity 
\beq
\sigma_{xy} ~=~ \frac{i}{\omega} ~\sum_{a \ne 0} ~ \Bigl[ ~\frac{\langle 0 
\vert J_x \vert a \rangle \langle a \vert J_y \vert 0 \rangle}{\omega - E_a 
+ E_0 + i \eta} ~-~ \frac{\langle 0 \vert J_y \vert a \rangle \langle a 
\vert J_x \vert 0 \rangle}{\omega + E_a - E_0 + i \eta} ~\Bigr] ~,
\label{sigj}
\eeq
where $\vert 0 \rangle$ is the ground state of the many-body system, and the 
sum over $\vert a \rangle$ runs over all the excited states; $\eta$ is an 
infinitesimal positive number. The current $\vec J$ is given by the 
second-quantized expression
\beq
{\vec J} ~=~ - ~c~ \frac{\delta {\cal H}}{\delta {\vec A}} ~=~ 
\frac{q}{2mi} ~ \sum_{\alpha=1}^2 ~\int d^2 {\vec r} ~ \Bigl[ ~
\Psi_{\alpha}^\dagger ~ \Bigl(~ {\vec p} - \frac{q_\alpha}{c} {\vec A} ~-~ 
\frac{q_\alpha}{c} {\vec {\cal A}}_{\alpha} ~) ~ \Psi_\alpha ~-~~ {\rm 
hermitian ~ conjugate} ~ \Bigr]~.
\label{curr}
\eeq
If we now perform the unitary transformation in (\ref{unit}) on both the 
current and the states, then (\ref{curr}) reduces to the conventional 
expression for the current operator of two species of fermions placed in the
magnetic fields given by (\ref{mag}). Eq. (\ref{sigj}) can then be evaluated 
in the usual way \cite{fett,chen}; in the zero frequency limit, we obtain the 
expression given in (\ref{sigxy}). The Hall conductivity will remain
unchanged if we make our model more realistic by including Coulomb repulsion
between the particles.

Another object of interest in this model is the matrix element of 
the ``hopping" operator 
\beq
M ({\vec r}) ~=~ c_1^\dagger ({\vec r}) ~c_2 ({\vec r})
\label{hop}
\eeq
between the ground state of the system with $(N_1 , N_2 )$ particles and all 
possible states of the system with $(N_1 +1, N_2 -1)$ particles. [The 
calculation of this overlap is of interest in connection with 
the ``orthogonality" catastrophe which is known to occur in Luttinger liquids 
in one dimension. It may also be useful in the context of a two-layer
quantum Hall system in which electrons can hop from one layer to the other].
Since our original Hamiltonian (\ref{ham}) is translation invariant, it is 
sufficient to compute the matrix element of $M ({\vec 0})$ located at the 
origin. This simplifies the computation for the following reason. In a second 
quantized form, the annihilation operator for any species is given by
\beq
c ({\vec r}) ~=~ \sum_{n,l} ~\psi_{n,l} ({\vec r}) ~c_{n,l} ~,
\eeq
where the sum runs over all one-particle states $(n,l)$ with wave functions 
$\psi_{n,l}$, and $c_{n,l}$ annihilates a fermion in the state $(n,l)$. Since 
only the zero angular momentum states have non-vanishing wave functions at the
origin, $c ({\vec 0})$ gets a contribution from only $l=0$ but all
possible radial quantum numbers $n$. Thus
\beq
c ({\vec 0}) ~=~ \sum_n ~\psi_{n,0} ({\vec 0}) ~c_{n,0} ~, 
\eeq
where 
\beq
\vert \psi_{n,0} ({\vec 0}) \vert^2 ~=~ \frac{qB}{2\pi \hbar c}
\label{lag}
\eeq
for {\it all} $n$. (This follows from the normalization of the Laguerre 
polynomials given in Refs. 7 and 8). We can now compute the 
frequency-dependent hopping function
\beq
{\cal M} (\omega ) ~=~ \sum_a ~\vert \langle a; N_1 + 1, N_2 -1  \vert ~M 
({\vec 0}) ~\vert 0; N_1 , N_2 \rangle \vert^2 ~2\pi ~\delta ( \hbar 
\omega ~-~ E_a ~+~ E_0 ) ~,
\label{overlap}
\eeq
where $\vert 0; N_1 , N_2 \rangle$ denotes the ground state of the system 
with $(N_1 , N_2 )$ particles, while $\vert a; N_1 +1, N_2 -1 \rangle$ denotes
all possible states of the system with $(N_1 +1, N_2 -1)$ particles. 

For simplicity, let us consider the case in which the filling fractions $f_1$ 
and $f_2$ are both less than $1$, and $N_1 = N_2 = N$. Then the 
ground state $\vert 
0; N , N \rangle$ is one which the both the type $1$ and type $2$ particles 
occupy the LLL states with $n=0$ and angular momentum $l=0,1,2,..., N -1$. 
Upon acting on this state with the operator $M ({\vec 0})$ in (\ref{hop}), we 
get a state $\vert a; N +1, N -1 \rangle$ in which a type $2$ particle has been
removed from the state $(0,0)$, and a type $1$ particle has been added to
the state $(n,0)$, where $n \ne 0$ due to the Pauli exclusion principle. Hence
the energy difference is 
\beq
E_a - E_0 ~=~ \frac{\hbar q}{mc} ~\Bigl[ ~(~n ~+~ \frac{1}{2} ~) ~\vert 
B_1 \vert ~-~ \frac{1}{2} ~\vert B_2 \vert ~\Bigr] ~,
\label{ea0}
\eeq
where $n=1,2,3,...$. This gives the locations of the $\delta$-functions on the
right hand side of (\ref{overlap}). We now have to find the weights. We can
use (\ref{lag}) to show that for species $2$,
\beq
\vert \langle a_2 ; N-1 \vert ~c_2 ({\vec 0}) ~\vert 0 ; N \rangle \vert^2 ~=~
\frac{q \vert B_2 \vert}{2\pi \hbar c} ~,
\label{over2}
\eeq
where $\vert a_2 ; N-1 \rangle$ represents the state in which a particle of 
type $2$ has been removed from the state $(0,0)$. Similarly, for species 
$1$, we have
\beq
\vert \langle a_1 ; N+1 \vert ~c_1^\dagger ({\vec 0}) ~\vert 0 ; N \rangle 
\vert^2 ~=~ \frac{q \vert B_1 \vert}{2\pi \hbar c} ~,
\label{over1}
\eeq
where $\vert a_1 ; N+1 \rangle$ represents the state in which a particle of 
type $1$ has been added to the state $(n,0)$ where $n \ge 1$. To use the 
results (\ref{over2}) and (\ref{over1}) for evaluating the matrix elements in 
(\ref{overlap}), we now perform the unitary transformation in 
(\ref{unit}). At this point, we have to worry about two things. Firstly, the 
$N+1$ particles of type $1$ in the states $\vert a \rangle$ see a slightly 
different magnetic field than the $N$ particles of type $1$ in the
state $\vert 0 \rangle$, since the number of
type $2$ particles differ by one in the two cases. However, the difference in
the two magnetic fields is of order $1/A$ which vanishes in the thermodynamic
limit. Thus the wave functions in the two cases look almost the same since the 
magnetic field $B$ appearing in (\ref{wave}) differs only slightly in the two 
cases; so when we use Eq. (\ref{over1}), it does not matter much if we set 
$N_2$ equal to $N$ or $N-1$ to determine the value of $B_1$ given by Eq.
(\ref{mag}). Secondly, we have to worry about the phase factor appearing in 
$U$ which depends on the total coordinates ${\vec R}_{N, \alpha} = \sum_i 
\rial$; recall Eq. (\ref{unit}). At this point, another advantage of 
locating the hopping operator at the origin ${\vec r} = {\vec 0}$ becomes 
apparent. Namely, we see that the phase factors ${\hat z} \cdot {\vec R}_{N+1,
1} \times {\vec R}_{N-1,2}$ appearing in the states $\vert a \rangle$ cancels 
with the phase factor ${\hat z} \cdot {\vec R}_{N,1} \times {\vec R}_{N,2}$ 
appearing in the state $\vert 0 \rangle$, since
the two states only differ by the addition or removal of particles at the
origin; this does not change the total coordinate ${\vec R}_\alpha$ of
either species. Putting Eqs. (\ref{over2}-\ref{over1}) and (\ref{ea0}) 
together, we see that the hopping function is given by
\beq
{\cal M} (\omega ) ~=~ \frac{q^2 \vert B_1 B_2 \vert}{(2\pi \hbar c)^2} ~
\sum_{n=1}^{\infty} ~2\pi ~\delta \Bigl( \hbar \omega ~-~ \frac{\hbar q}{mc} 
(n ~ +~ \frac{1}{2}) ~\vert B_1 \vert ~+~ \frac{\hbar q}{2mc} ~\vert B_2 
\vert ~\Bigr) ~.
\eeq
We thus get an infinite sequence of $\delta$-functions with equal weight.

Finally, let us compute the one-particle off-diagonal density matrix for, 
say, species $1$. We assume again that $N_1 = N_2 = N$ and both the filling 
fractions $f_\alpha$ are less than $1$. We have to evaluate
\beq
\rho ({\vec r}, {\vec r}^\prime ) = \int ~\prod_{i=2}^N d^2 {\vec r}_{
i,1} \prod_{j=1}^N d^2 {\vec r}_{j,2} ~
\psi^\star ({\vec r}, {\vec r}_{2,1}, ..., {\vec r}_{N,1}; {\vec r}_{1,2}, 
{\vec r}_{2,2}, ..., {\vec r}_{N,2}) ~\psi ({\vec r}^\prime, {\vec 
r}_{2,1}, ..., {\vec r}_{N,1}; {\vec r}_{1,2}, {\vec r}_{2,2}, ..., {\vec 
r}_{N,2}) ,
\label{off}
\eeq
where we assume that the particles fill up the states $l=0,1,2,...,N$ in the
LLL. We again perform the unitary transformation (\ref{unit}). The integrand
in (\ref{off}) then becomes the product of a phase factor
\beq
\exp ~[~\frac{iq\hbar \gamma}{cA} {\hat z} \cdot ({\vec r} - {\vec r}^\prime )
\times {\vec R}_2 ~] ~
\eeq
(where ${\vec R}_2 = \sum_j {\vec r}_{j,2}$ and we have used Eq. (\ref{gam})), 
four Van der Monde determinants which typically look like $\prod_{k<l} 
(z_{k, \alpha } - z_{l, \alpha })$ and its complex conjugate
for both the species, and the Gaussian factor
\beq
\exp ~[~ - \frac{q \vert B_1 \vert}{4\hbar c} ({\vec r}^2 + {\vec r}^{\prime
2} + 2 \sum_{i=2}^N {\vec r}_{i,1}^2 ) ~-~ \frac{q \vert B_2 \vert}{2\hbar 
c} \sum_{j=1}^N {\vec r}_{j,2}^2 ~] ~.
\eeq
Since the Van der Monde determinants are invariant under translations, we can
immediately integrate over the $N-1$ independent relative coordinates (i.e., 
${\vec r}_{k,2} - {\vec r}_{l,2}$) of the type $2$ particles. 
The total coordinate ${\vec R}_2$ of species $2$ then remains in the form
\beq
\exp ~[~ \frac{iq\hbar \gamma}{cA} {\hat z} \cdot ({\vec r} - {\vec r}^\prime )
\times {\vec R}_2 ~-~ \frac{q \vert B_2 \vert}{2\hbar c} 
\frac{{\vec R}_2^2}{N} ~]~.
\eeq
When we integrate over ${\vec R}_2$, we get a Gaussian of the form
$\exp [- d ({\vec r} - {\vec r}^\prime )^2 /A]$, where $d$ is a number of
order $1$. In the thermodynamic limit, we can set this equal to $1$ since
we can assume that the separation $\vert {\vec r} - {\vec r}^\prime \vert$ is
much smaller than the size of the system. We are now left with only the
coordinates ${\vec r}_{i,1}$, with $i=2,3,...,N$, to integrate over. We 
finally get
\beq
\rho ({\vec r}, {\vec r}^\prime ) ~=~ \frac{1}{N} ~\sum_{l=0}^N ~\Bigl( ~
\frac{qB_1}{2\hbar c} ~\Bigr)^{l+1} ~\frac{(z z^{\prime \star} )^l}{l! ~\pi} ~
\exp ~[~-~ \frac{qB_1}{4\hbar c} (z z^\star ~+~ z^\prime z^{\prime 
\star}) ~]~. 
\eeq
where we have assumed that $qB_1$ is positive. If we now take the limit
$N \rightarrow \infty$, we find that the off-diagonal density matrix is
the product of a Gaussian times a phase,
\beq
\rho ({\vec r}, {\vec r}^\prime ) ~\sim ~ \frac{qB_1}{2\pi \hbar c} ~
\exp ~\Bigl[~-~ \frac{qB_1}{4\hbar c} \vert z ~-~ z^\prime \vert^2 ~-~ 
\frac{qB_1}{4\hbar c} (z^\star z^\prime ~-~ z z^{\prime \star} ) ~\Bigr] ~,
\eeq
which is the usual result for a single species of particles in the LLL.

To summarize, we have introduced and solved a two-dimensional multi-species
fermi system with mutual interactions of a particular type. The interaction 
can be converted via an unitary transformation into a static magnetic field 
whose strength depends on the density of particles. The model is quite simple;
after all, the exact solvability of a Hamiltonian which has a quadratic form
should surprise no one. Yet the physics of the model is quite interesting. We
end up with a strongly non-fermi liquid system; further, the elements of the
orthogonality catastrophe, i.e., a readjusting of {\it all} states in response
to the addition of a single particle, also carry over from the one-dimensional 
physics of the Luttinger model. Our model may thus serve some purpose in 
understanding the physics of non-fermi liquids in higher dimensions.

We would like to dedicate this paper to the memory of Heinz Schulz.


\end{document}